\def\year{2022}\relax
\documentclass[letterpaper]{article} 
\usepackage{aaai22}  
\usepackage{times}  
\usepackage{helvet}  
\usepackage{courier}  
\usepackage[hyphens]{url}  
\usepackage{graphicx} 
\usepackage{natbib}  
\usepackage{caption} 
\DeclareCaptionStyle{ruled}{labelfont=normalfont,labelsep=colon,strut=off} 
\usepackage{hyperref} 

\usepackage{amsmath}
\usepackage{amsthm}
\usepackage{amssymb}
\usepackage{enumitem}
\usepackage{microtype}
\usepackage{subfigure}
\usepackage{booktabs} 
\usepackage{algorithm}
\usepackage{algorithmic}
\usepackage{xspace}
\usepackage{bm}
\usepackage{cite}
\usepackage{booktabs}  
\usepackage{threeparttable}  
\usepackage{multirow}
\usepackage{multicol}
\usepackage{arydshln}
\usepackage{color}

\theoremstyle{plain}

\theoremstyle{definition}

\usepackage{newfloat}
\usepackage{listings}
\floatstyle{ruled}
\newfloat{listing}{tb}{lst}{}
\floatname{listing}{Listing}

\title{\scheme: A Blockchain-based \\Secure Data Market for Decentralized Machine Learning}

\title{\scheme: A Blockchain-based \\Secure Data Market for Decentralized Machine Learning}
\author {
    Jiacheng Liang\textsuperscript{\rm 1,2}
    Songze Li \textsuperscript{\rm 1}
    Bochuan Cao \textsuperscript{\rm 2}
    Wensi Jiang \textsuperscript{\rm 1}
    Chaoyang He \textsuperscript{\rm 3}
}

\affiliations {
    \textsuperscript{\rm 1} Hong Kong University of Science and Technology\\
    \textsuperscript{\rm 2}University of Electronic Science and Technology of China\\
    \textsuperscript{\rm 3} University of Southern California\\
    jliangbb@connect.ust.hk, songzeli@ust.hk,  bochuancao@outlook.com, wensi.jiang@connect.ust.hk, chaoyang.he@usc.edu
}

\newcommand{\scheme}{{\sf OmniLytics}\xspace}
\newcommand{\MO}{{\sf MO}\xspace}
\newcommand{\DO}{{\sf DO}\xspace}
\newcommand{\contract}{{\sf SecModelUpdate}\xspace}
\newcommand{\setup}{{\sf Setup}\xspace}
\newcommand{\register}{{\sf Register}\xspace}
\newcommand{\GA}{{\sf ModelAggregate}\xspace}
\newcommand{\OS}{{\sf OutlierSuppression}\xspace}
\newcommand{\payment}{{\sf Payment}\xspace}
\newcommand{\finish}{{\sf Finished}\xspace}
\newcommand{\whitelist}{{\sf whitelist()}\xspace}
\newcommand{\start}{{\sf start()}\xspace}
\newcommand{\exit}{{\sf exit()}\xspace}
\newcommand{\PI}{{\sf PubKeyInteract}\xspace}

\renewcommand{\eth}{{\sf Ethereum}\xspace}

\renewcommand{\P}{{\mathcal{P}}}
\renewcommand{\L}{{\cal L}}
\renewcommand{\S}{{\cal S}}
\newcommand{\vct}[1]{\bm{#1}}

\DeclareMathOperator*{\argmin}{arg\,min}

\begin{document}

\maketitle

\begin{abstract}
We propose \scheme, a blockchain-based secure data trading marketplace for machine learning applications. Utilizing \scheme, many distributed data owners can contribute their private data to collectively train an ML model requested by some model owners, and receive compensation for data contribution. \scheme enables such model training while simultaneously providing 1) model security against curious data owners; 2) data security against the curious model and data owners; 3) resilience to malicious data owners who provide faulty results to poison model training; and 4) resilience to malicious model owners who intend to evade payment. \scheme is implemented as a blockchain smart contract to guarantee the atomicity of payment. In \scheme, a model owner splits its model into the private and public parts and publishes the public part on the contract. Through the execution of the contract, the participating data owners securely aggregate their locally trained models to update the model owner's public model and receive reimbursement through the contract. We implement a working prototype of \scheme on \eth blockchain and perform extensive experiments to measure its gas cost, execution time, and model quality under various parameter combinations.
For training a CNN on the MNIST dataset, the $\MO$ is able to boost its model accuracy from 62\% to 83\% within 500ms in blockchain processing time.
This demonstrates the effectiveness of \scheme for practical deployment.
\end{abstract}
\section{Introduction}
With the rapid development of sensing, processing, and storage capabilities of computing devices (e.g., smartphones and IoT devices), the collection and storage of data has been increasingly convenient and cost-effective~\cite{sheng2013sensing,cornet2018systematic}. On the other hand, crowdsourcing big data has been shown to be extremely effective in improving the performance of various fields such as healthcare, smart city, and recommender systems~\cite{raghupathi2014big,wang2018big,hashem2016role,al2015applications,yin2013lcars}. The abundant supply of data stored locally at individual nodes and the large demands from data-intensive applications incentivise the development of a data market on which data owners can easily trade the rights of using their data with interested consumers for monetary returns.

Conventionally, a data market is often operated as a centralized service platform that collects data from data owners and sells raw or processed data to the consumers~\cite{mivsura2016data,krishnamachari2018i3,niu2018achieving}. This approach leaves the platform as a single point of security vulnerability for the data market, and corruption on the platform servers may lead to severe security issues including leakage of private data, faulty computation results, and manipulation of data price. A number of recent works have proposed to leverage technologies of decentralized systems like blockchains and smart contracts to tackle the weakness of the centralized implementation (see, e.g.,~\cite{ozyilmaz2018idmob,duan2019aggregating,ramachandran2018towards,koutsos2020agora}). 

To further improve data security, more advanced cryptographic techniques like homomorphic and functional encryption have been utilized to generate analytics over the raw data for consumers to purchase without revealing the data themselves~\cite{duan2019aggregating,niu2018unlocking,niu2018achieving,koutsos2020agora}. However, these approaches are limited in the following three aspects: 1) data owners upload the encrypted raw data on the blockchain, which leads to permanent loss of the data ownership to any adversarial party with the decryption key; 2) the available analytics are limited to simple operations like linear combinations; 3) they still require a (trusted or untrusted) third party other than the data owner and consumer acting like a broker or service provider to maintain the utility and security of trading session. 


In this paper, our goal is to build a \emph{secure} and \emph{broker-free} data market for \emph{general machine learning} applications. Particularly, a model owner would like to crowdsource training data from interested data owners through the data market for improving the quality of its ML models (e.g., a deep neural network for image classification). Moreover, we enforce the following privacy and security requirements:
\emph{Model privacy}: parameters of the trained ML model are (mostly) kept private to the model owner itself and are not revealed to other parties; \emph{Data privacy}: a model owner learns nothing about data owners' private data other than the learnt model; \emph{Byzantine resistance}: 1) robust to malicious data owners who intentionally provide incorrect computation results, and 2) robust to malicious model owners who try to evade payment.

\begin{figure}[htbp]
   \centering
   \includegraphics[width=\linewidth]{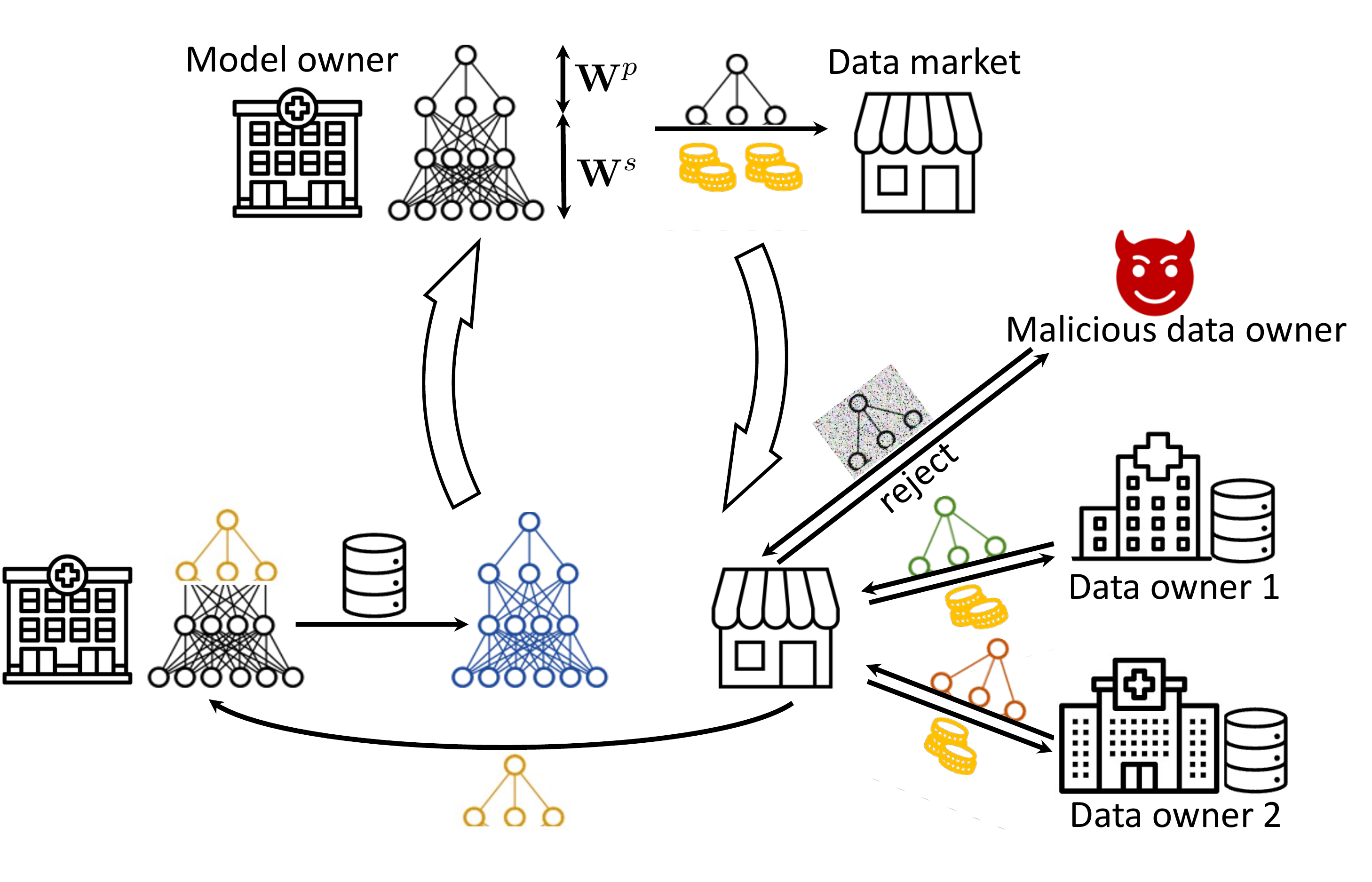} 
   \vspace{-8mm}
   \caption{An overview of the operation of the proposed \scheme data market. The model owner splits its initial model into a private model $\vct{W}^s$ and a public model $\vct{W}^p$, updates the public model through the data market using data owners' private data, and combines the updated public model with the old private model and performs local model adaptation with its private data. The smart contract implementing the data market reimburses the honest data owners who contribute to updating public model, and rejects erroneous results from malicious data owners.}
   \label{fig:overview}
\vspace{-0.6cm}
 \end{figure}

Our main contribution is the construction of a novel blockchain-based data market named \scheme, which is the first implementation of an \eth smart contract~\cite{ETH} that simultaneously satisfies all the above security requirements. As shown in Figure~\ref{fig:overview}, during a trading session, the model owner splits its initial model into a private part that contains most of the model parameters, and a public part with a much smaller size, and
deploys a smart contract only containing the public model. This model splitting helps to protect the privacy of model parameters, and to reduce the computation complexity hence the gas cost of running the contract. Upon observing the contract, interested data owners retrieve the public model, update the model locally using their private data, and upload the updated models back to the contract for further aggregation. During this process, the data owners hide their local models with pair-wise masks generated according to the secure aggregation protocol~\cite{bonawitz2017practical} to further protect from data leakage. Moreover, \scheme implements the multi-Krum algorithm~\cite{blanchard2017machine,blanchard2017byzantine} to remove faulty computation results from malicious data owners. Finally, the model owner fetches the updated public model, concatenates it with its private model, and continues to train the combined model with the private data. The training reward is provided by the model owner when deploying the contract and is automatically distributed to honest data owners by the contract, preventing malicious model owners from evading payments after obtaining the trained model.

We implement an \eth smart contract \contract and the off-chain application of the proposed \scheme data market. We conduct extensive experiments to measure the gas cost, execution time of \contract, and trained model accuracy under various combinations of system and design parameters. 
For instance, for a training task on the MNIST dataset, the $\MO$ is able to boost its model accuracy from 62\% to 83\% using \scheme, with less than 500ms in blockchain processing time.

\subsection*{Related works}
{\bf Secure data markets.}
Traditional data markets require full trust on the centralized service platform, which leaves the platform a single point of failure. To resolve this issue, implementing data market over decentralized systems like blockchains has been recently proposed~\cite{ozyilmaz2018idmob,ramachandran2018towards,banerjee2018blockchain,duan2019aggregating,koutsos2020agora}. In these implementations, encrypted data are uploaded to the blockchain, on which a smart contract with a funding deposit from the consumer is executed automatically, guaranteeing the atomicity of the payment. To further enhance the data privacy and robustness against malicious behaviors, more advanced techniques like homomorphic encryption, functional encryption, and differential privacy have been utilized to securely generate simple analytics over the raw data for sale~\cite{duan2019aggregating,niu2018unlocking,koutsos2020agora}, and zero-knowledge proofs and trusted hardware like Intel SGX have been used to guarantee the correctness of the computations~\cite{duan2019aggregating,niu2018achieving,koutsos2020agora}. 

\noindent {\bf Federated learning on blockchains.} Federated learning (FL)~\cite{mcmahan2017communication} has recently emerged as a privacy-preserving framework for distributed ML, where a set of clients, instead of directly uploading their private data to the cloud, upload the gradients computed from the data, which are aggregated at a cloud server to update a global model. In addition, techniques of masking local gradients like secure aggregation~\cite{bonawitz2017practical,bell2020secure,so2021turbo} and differential privacy~\cite{geyer2017differentially,wei2020federated} have been developed for FL to further protect 
data privacy.

Recently, it has been proposed to execute FL tasks on blockchains to combat server corruption and facilitate a more fair and transparent reward distribution (see, e.g.,~\cite{zhao2020privacy,liu2020fedcoin,kim2019blockchained,shayan2020biscotti,ma2020federated}). In~\cite{zhao2020privacy}, an FL on blockchain system is designed for the smart appliance manufactures to learn a ML model from customers' data. Differential privacy techniques are applied to protect data privacy. 
One major weakness of the design in~\cite{zhao2020privacy} is that the learnt model of the manufacturer is completely revealed to public, which may not be desirable for the model owner who pays for training. 
In~\cite{lyu2020towards}, an FL framework over blockchain named FPPDL was proposed to facilitate fair and privacy-preserving machine learning. While both FPPDL and the proposed \scheme data market protect data privacy via encrypting model updates,
they differ drastically in the scope of applicability. In FPPDL, a data owner is co-located with a blockchain miner such that a learning participant has to perform local training and block verification.
This restricts FPPDL to permissioned blockchains. In contrast, \scheme is designed as a plug-and-play smart contract on top of any blockchain platform, without needing to modify miners' operations. This allows \scheme to be easily deployed on permissionless public blockchains like \eth.
\section{Secure Data Market for Decentralized Machine Learning}
We consider a network of many compute nodes (e.g., mobile devices like smartphones, or institutions like hospitals, banks, and companies), each of which has some local data storage and processing power. Some nodes would like to obtain a machine learning model (e.g., to predict certain disease from patients' exam data). We call such node the model owner, denoted by $\MO$. However, as the local data possessed by $\MO$ may not be sufficient to train a good model with high accuracy, the $\MO$ intends to crowdsource data from other nodes to improve the model quality. In return, the $\MO$ compensates the nodes who contribute to the model training with their local private data. We call these nodes as data owners, denoted by $\DO$s.

We consider a threat model where neither the $\MO$ nor the $\DO$s can be trusted, i.e., they may attempt to recover the private data of other participants, deviate from the agreed data trading protocol, or intentionally evade payments. The goal of this paper is to design a secure data market that meets the following specific security requirements:

\begin{itemize}[leftmargin=*]
    
    \item {\bf Model privacy.} The $\MO$ would like to keep its
    (complete) model secret from the $\DO$s and the data market, because 1) the $\MO$ pays to train the model, which may provide it with competitive advantage over other parties; and 2) the model itself may be pre-trained using $\MO$'s private data that contains confidential information about $\MO$.
    
    \item {\bf Data privacy.} A $\DO$ would like to keep its private data secret from the $\MO$ and other $\DO$s. 
    
    \item{\bf Resistance to Byzantine data owners.} The data market should protect the quality of the trained model from being undermined by malicious $\DO$s, who might arbitrarily deviate from the trading protocol, and supply faulty results. 
    
     \item{\bf Resistance to Byzantine model owner.} The data market should restrain malicious $\MO$ from escaping the payment, and enforce that honest $\DO$s who faithfully follow the protocol are properly compensated.
\end{itemize}

\vspace{-0.2cm}
\section{Secure Federated Machine Learning}\label{sec:SFL}
In this section we review the federated learning (FL) framework~\cite{mcmahan2017communication} and some of its security-enhancing techniques, which constitute the starting point towards our design of a secure data market.
 

\subsection{Federated learning framework}
A FL network consists of a central server and a group of $N$ clients. Each client $k$ locally has a dataset $\S_k =\{(\vct{x}_1,\vct{y}_1),\ldots,(\vct{x}_{M_k},\vct{y}_{M_k})\}$ of $M_k$ data samples. Each data sample $(\vct{x}_i,\vct{y}_i)$ consists of an input vector $\vct{x}_i \in \mathbb{R}^d$ and its label $\vct{y}_i \in \mathbb{R}^p$ for some input dimension $d$ and output dimension $p$. The server aims to train a global model $\vct{W}$ (e.g., a deep neural network) to minimize the objective function $\mathcal{L}(\vct{W}) = \sum_{k=1}^N p_k \L_k(\vct{W})$.
Here $\L_k(\vct{W}) = \frac{1}{M_k} \sum_{i=1}^{M_k} \ell(\vct{W};(\vct{x}_i,\vct{y}_i))$ is the empirical loss at client $k$ for some loss function $\ell$. The weight $p_k \triangleq \frac{M_k}{\sum_{k=1}^N M_k}$.

The server collaborates with the clients to train the global model using the {\sf FedAvg} algorithm~\cite{mcmahan2017communication} over multiple iterations. To start with, the server broadcasts an initial model $\vct{W}^{(0)}$ to all clients. In iteration $t$, each client $k$ first splits its local dataset $\S_k$ into batches of size $B$, and starting from the global model received in last iteration $\vct{W}^{(t-1)}$, runs for $E$ local epochs batched-gradient descent through $\S_k$ to obtain local model $\vct{W}_k^{(t)}$, and sends it to the server. The server aggregates the received gradients from the $N$ clients and updates the global model as $\vct{W}^{(t)} = \sum_{k=1}^N p_k \vct{W}_k^{(t)}$.
\vspace{-0.2cm}
\subsection{Secure model aggregation}\label{sec:secure_agg}
While the FL framework was designed to protect the privacy of clients' data by having them send merely the models (rather than the actual data) to the server, it was shown in ~\cite{zhu2020deep,wang2019beyond,geiping2020inverting} that the server can recover private data ${\cal S}_k$ from the local model $\vct{W}_k^{(t)}$ and the global model $\vct{W}^{(t-1)}$ via model inversion attacks. To prevent such data leakage, the secure model aggregation protocol was proposed in~\cite{bonawitz2017practical} to mask local models before uploading to the server. Specifically, each client $k$ sends a masked model $\widetilde{\vct{W}}_k = \vct{W}_k + \vct{Z}_k$ with some random mask $\vct{Z}_k$ to the server. 
The secure aggregation protocol guarantees 1) server cannot deduce any information about each individual $\vct{W}_k$ and hence the private data $\S_k$; and 2) the masks are generated such that $\sum_{k=1}^N \vct{Z}_k = \vct{0}$, and server can exactly recover the aggregation of the local model weights $\sum_{k=1}^N\widetilde{\vct{W}}_k = \sum_{k=1}^N\vct{W}_k$. 

\vspace{-0.2cm}
\subsection{Byzantine-resilient federated learning}
In FL, malicious clients can manipulate the models uploaded to the server to poison the global model~(see, e.g.,~\cite{bhagoji2019analyzing,fang2020local}), or plant backdoors in the global model (see, e.g.,~\cite{bagdasaryan2020backdoor,NEURIPS2020_backdoor}). Current strategies to defend Byzantine clients mainly follow distributed machine learning protocols designed under adversarial settings~\cite{blanchard2017machine,chen2017distributed,yin2018byzantine,yang2019byzantine,li2019rsa,ghosh2020communication,data2021byzantine}. The main idea is to take advantage of statistical similarities among models from honest clients, and remove or suppress the adversarial effects introduced by faulty results from Byzantine clients. For instance the Krum algorithm, proposed in~\cite{blanchard2017machine}, detects the gradient vectors that have large $\ell_2$-distances from the others as being from adversarial clients, and removes them from the aggregation process.

\section{The Proposed Secure Data Market}
While the above techniques can help to protect $\DO$s' data privacy, and defend against Byzantine $\DO$s in the federated learning framework, they are insufficient to meet our security requirements of keeping model secret from the $\DO$s, and robustness against malicious $\MO$, for a data market.

We propose to resolve these issues via adopting a private-public model splitting paradigm at $\MO$, and leveraging blockchain technologies to securely collect and reimburse $\DO$s' contributions to train the public model. Specifically, we design \scheme, an end-to-end solution to provide a transparent, fair, yet private and secure data market. In the rest of the section we describe in detail the \scheme data market, which includes the local computations of the $\MO$ and $\DO$s, the design of a smart contract implementing secure model update and reward distribution, and the interactions between the model and data owners with the contract.

\vspace{-0.2cm}
\subsection{Private-public model splitting}
We consider the scenario where the $\MO$ starts with some initial model $\vct{W}$. In practice, this is often obtained by the $\MO$ through adapting a pre-trained foundation model (e.g., BERT for language understanding or DALL-E for image generation) to its local task and data~\cite{bommasani2021opportunities}. 


To improve the accuracy of the initial model $\vct{W}$, the $\MO$ pays the $\DO$s to further refine the model using their local private data. Following the local-global model splitting paradigm for federated learning~\cite{liang2020think}, the $\MO$ first splits the initial model $\vct{W} = (\vct{W}^s, \vct{W}^p)$ into a private model $\vct{W}^s$ and a public model $\vct{W}^p$. The splitting is performed such that the public model is much smaller than the private one. For instance, the public model could be the last few layers of a large DNN. After the splitting, the $\MO$ submits the public model $\vct{W}^p$ to the data market to be updated by the participating $\DO$s using their private data. Once the public model is updated over the data market and returned to the $\MO$, the $\MO$ concatenates the updated public model and the old private model, and performs another round of model adaptation using local data, improving the model accuracy. Subsequently, as shown in Figure~\ref{fig:overview}, the $\MO$ can repeat this pipeline where public model is first updated via the data market, and can then be combined with the private model for local adaptation until no significant improvement is observed on model accuracy. 

The approach of splitting $\MO$'s model into private and public parts provides the following salient benefits:
\begin{itemize}[leftmargin=*]
    \item \emph{Model privacy}: most model parameters of the $\MO$ are contained in the private model $\vct{W}^s$, and kept secret from the data market and the $\DO$s;
    \item \emph{Computation and communication complexities}: as only a small number of model parameters in the public model $\vct{W}^p$ need to be updated and communicated over the data market, the computation and communication costs at the market and the $\DO$s are significantly smaller, compared with operating over the entire model.
\end{itemize}

We next describe our design of a smart contract to execute the secure update of the public model $\vct{W}^p$. Since the contract exclusively deals with the public model, for ease of exposition, we will drop the superscript of $\vct{W}^p$, and simply refer to the public model as the model. 

\subsection{Smart contract design for secure model update}
The proposed \scheme data market implements the model update and reward distribution through a smart contract named \contract, over an underlying blockchain (e.g., \eth). The $\MO$ and the $\DO$s are users with dedicated addresses who submit transactions to change the state of \contract, while the actual execution of the contract program and the recording of the contract state are performed by blockchain miners and verifiers, who may or may not co-locate with the $\MO$ and $\DO$s.

\contract executes a single iteration of the model update, with the initial model from the $\MO$, and the model updates collected up to as many as $R$ rounds and from up to $N$ $\DO$s in each round. The use of smart contract enforces automatic payment towards participating $\DO$s whose computation results are considered valid, via some verification mechanism implemented on the contract. 

\begin{figure}[htbp]
   \centering
   \includegraphics[width=\linewidth]{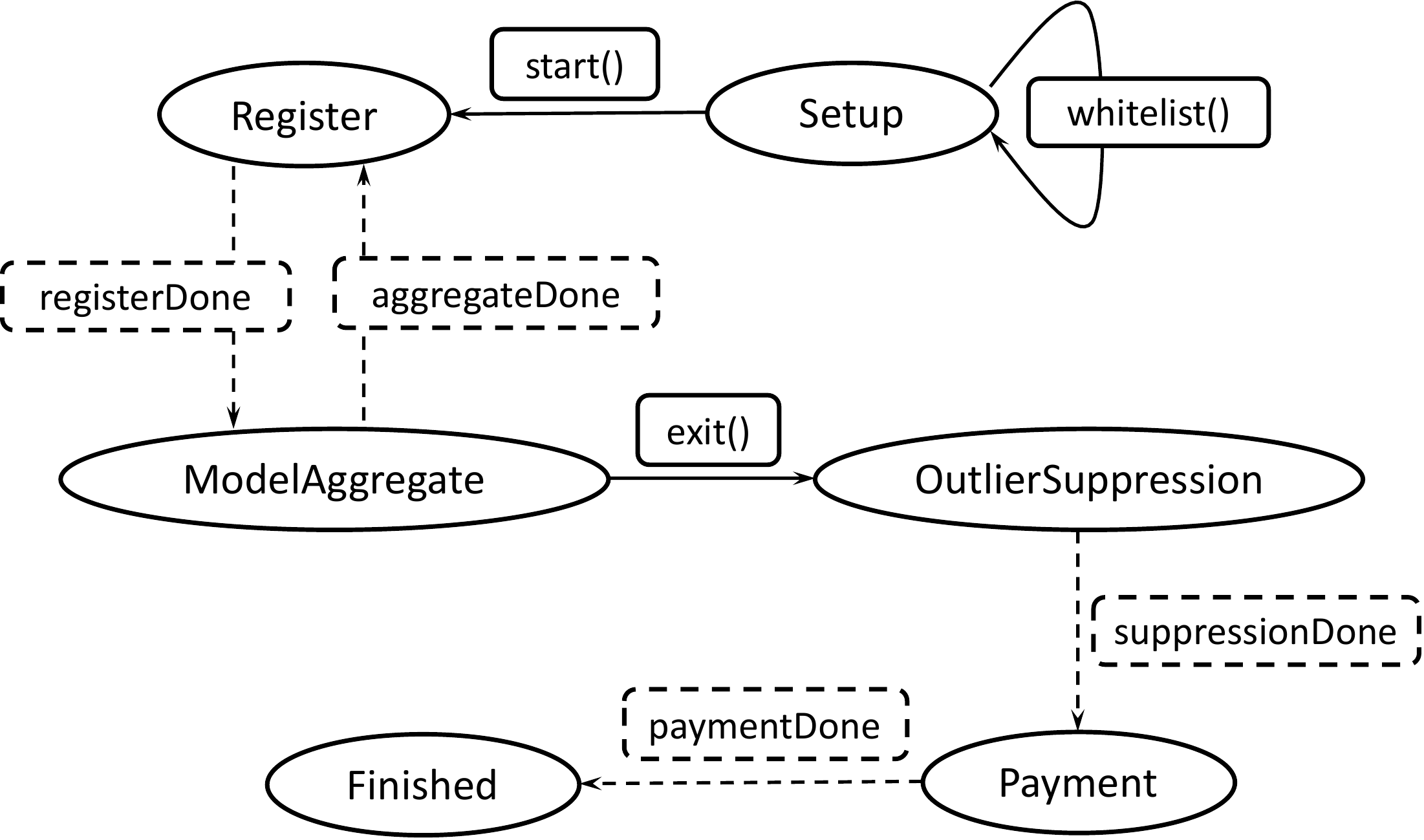} 
   \caption{State transition of the smart contract \contract. The six states are represented by ovals. State transitions are triggered by either applying a method (in a solid box), or occurrence of an event (in a dashed box).}
   \label{fig:states}
\vspace{-0.6cm}
 \end{figure}

\subsubsection{Model collection and aggregation}
The $\MO$ initializes the data trading session by deploying a smart contract \contract with a training reward deposit on the blockchain. As shown in Figure~\ref{fig:states}, \contract transitions between six states, i.e., \setup, \register, \GA, \OS, \payment, and \finish. Upon deployment, \contract is in the \setup state with a set of $\DO$s the $\MO$ would like to purchase data from specified by a \whitelist method. The $\MO$ issues a transaction with the \start method specifying the following public parameters:
\begin{itemize}[leftmargin=*]
    \item The initial model parameters $\vct{W}$;
    \item Minimum number of data points required for each participating $\DO$ to compute its local model, denoted by $M_0$;
    \item Minimum number of local epochs $E$; 
    \item Local batch size $B$;
    \item Number of rounds to collect models, denoted by $R$;
    \item Maximum number of distinct $\DO$s to collect gradients from within each round, denoted by $N$.
\end{itemize}

Executing \start moves \contract into the \register state, and the contract starts to register for the $\DO$s who intend to participate in the model aggregation for the first round. In each round $r$, $r=1,\ldots,R$, after $N$ $\DO$s have registered for this round, \contract moves to the \GA state and starts to collect local computation results from the $\DO$s. For each $k=1,\ldots,N$, the $k$th $\DO$ registered for round $r$ reads $\vct{W}$ from the contract, and performs local training using $M_0$ private data points, as specified in the previous section, obtaining the local model $\vct{W}_{r,k}$.
Next, the $\DO$s in round $r$ execute the secure aggregation protocol to generate the random masks $\vct{Z}_{r,1},\ldots,\vct{Z}_{r,N}$, and the $k$th $\DO$ sends the masked model $\widetilde{\vct{W}}_{r,k} = \vct{W}_{r,k} + \vct{Z}_{r,k}$ to the contract, for $k=1,\ldots,N$.

After receiving the masked models from all registered $\DO$s in round $r$, the contract aggregates them to obtain 
\begin{align}
\vct{A}_r =\frac{1}{N}\sum_{k =1}^{N} \widetilde{\vct{W}}_{r,k}= \frac{1}{N}\sum_{k =1}^{N} \vct{W}_{r,k}.
\end{align}
If the results of some $\DO$s are still missing after certain amount of time, the aggregation in round $r$ fails and we have $\vct{A}_r = \emptyset$. 

After the aggregation process of round $r$, \contract moves back to the \register state for the next round $r+1$. Only $\DO$s whose local computation results have not been incorporated in the aggregation results of any previous rounds are eligible to register. By the end of the aggregation process the contract \contract obtains a set of results $\{\vct{A}_r\}_{r \in \P}$, where $\P \subseteq \{1,\ldots,R\}$ denotes the indices of the rounds in which the secure aggregation had been successfully executed. The contract transits to the \OS state if $R$ rounds of gradient aggregation have been executed or is manually triggered by the \exit method.

\subsubsection{Outlier Suppression}
\label{OS_label}
During the model collection and aggregation process, Byzantine $\DO$s may upload maliciously generated models which corrupt the aggregation results in $\{\vct{A}_r\}_{r \in \P}$. \contract adopts Byzantine resistance techniques to remove outliers in the \OS state. Specifically, we assume that at most $\mu < \frac{1}{2}$ fraction of the aggregation results $\{\vct{A}_r\}_{r \in \P}$ may be corrupted by malicious $\DO$s. The contract runs a distance-based outlier suppression algorithm $m$-Krum~\cite{blanchard2017machine,blanchard2017byzantine} (see Algorithm~\ref{alg:MK} in Appendix A) on $\{\vct{A}_r\}_{r \in \P}$ to select a subset $\P' \subset \P$ of $|\P'|=m$ aggregation results that are considered computed correctly, for some $m < (1-2\mu)|\P|-2$.

\subsubsection{Reward distribution}
The contract enters the \payment state after the outlier suppression and the set $\P'$ is obtained. The training reward deposited by the $\MO$ on the contract is evenly distributed into the accounts of the $\DO$s, whose computation results from rounds in $\P'$ have been accepted.

\subsubsection{Public model update}
After the execution of the \contract contract, the $\MO$ obtains from the contract the selected aggregation results $\{\vct{A}_r\}_{r \in \P'}$, and updates its public model to
\begin{equation}\label{eq:final_grad}
    \vct{W}=\frac{1}{|\P'|}\sum_{r \in \P'} \vct{A}_r.
\end{equation}


\section{Security Analysis}
In this section, we analyze the security properties of \scheme. Particularly, our analysis includes the following four aspects: 1) the confidentiality of the $\MO$'s model parameters; 2) the privacy of each $\DO$'s data; 3) the security of the model update; and 4) the correct execution of the \contract contract. 

\subsection{Confidentiality of model parameters}
During the execution of the \contract contract, only $\MO$'s public model containing a small number of parameters is revealed to the data market and the $\DO$s, while the majority of the model parameters in the private model are kept secret to the $\MO$ itself. 

\subsection{Confidentiality of local data}
While each data owner can participate in model aggregation in at most one round in the contract \contract, its private data is only related to the aggregation result $\vct{A}_r$ for a single round $r$. We consider the situation where a subset ${\cal C} \subset \{1,\ldots,N\}$ of $\DO$s in round $r$ may collude to infer the private data of some $\DO$ in the same round. Based on the privacy guarantee of the secure aggregation protocol~\cite{bonawitz2017practical} employed by the contract, we argue that as long as the number of colluding $\DO$s $|{\cal C}|$ is less than some secure parameter $T$\footnote{Each $\DO$'s secret keys to generate random masks are secret shared with other $\DO$s such that any $T$ colluding $\DO$s can reveal the secret keys of any data owner.}, no information about other $\DO$s' private data other than the aggregation result $\vct{A}_r$ can be inferred. 





\subsection{Security of model update}
To combat malicious data owners uploading faulty computation results to the contract, we employ the $m$-Krum algorithm from~\cite{blanchard2017byzantine} to select the aggregation results from a subset of $\P' \subset \P$, which are considered to be close to the expected value with respect to the underlying data distribution. 


We consider the scenario where all $\DO$s have i.i.d. data. In this case, all the aggregation results $\{\vct{A}_r\}_{r \in \P}$ from the successful rounds of \contract are independently and identically distributed. This is because each honest data owner follows the same local training operation of {\sf FedAvg}~\cite{mcmahan2017communication} with $M_0$ data points, and each round aggregates results from $N$ $\DO$s. Therefore, according to Proposition 2 and Proposition 3 in~\cite{blanchard2017byzantine}, as long as $|\P'| < (1-2\mu)|\P|-2$, where $\mu$ is the maximum fraction of the aggregation results that may be corrupted, the estimated overall gradient in~(\ref{eq:final_grad}) provides a close approximation of the true gradient, which leads to the convergence of the model training. 

\section{Ethereum Implementation and\\ Experimental Results}
We implement a working prototype of \scheme over \eth, using Solidity \cite{Solidity} to develop the contracts, Python for the off-chain applications and PyTorch \cite{pytorch} for the neural network training. We deploy the smart contract \contract via the Remix IDE \cite{Remix} on the local Geth Ethereum Testnet \cite{geth}. We connect the off-chain applications to the smart contracts using the Web3py library \cite{Web3py} and monitor the created transactions using Etherscan. Each data owner is connected to the Geth network with a unique Ethereum address. We conduct experiments on a machine with AMD R5-5600X CPU @3.70 GHz, Nvidia RTX3070 GPU, 32 GB of Memory and 1 TB SSD. 

\contract consists of four major operations: \register, \PI, \GA, and \OS. Once a \contract contract is deployed on \eth by a model owner, data owners greedily register with the contract to upload their local computation results until the results are incorporated in the final model aggregation. Data owners within the same aggregation group runs \PI to exchange public keys for secure aggregation as done in~\cite{bonawitz2017practical}. Data owners pay for the transaction fee to upload the results and secure aggregation, which will be reimbursed by the model owner in the \payment phase.

During the secure aggregation process in each round, while the default Pytorch data type is $float32$, we scale each value by $10^8$ and aggregate the integer part to improve the precision of the aggregation result.
We turn on automatic mining mode and set the mining time to generate a new block to 1 second.

\noindent {\bf Parallel group aggregation.} We perform a system-level optimization such that each round of secure aggregation is carried out in parallel to speed up contract processing. This means that if all the data owners in one round have submitted their local results, the process of secure aggregation would be performed. There is no need to wait until previous rounds are completed. When the last round is completed, model owner can initiate the multi-Krum process to obtain the final result. 

Aside from experimental results reported in this section, we provide additional results in Appendix B.

\subsection{Smart contract measurements}
We first measure the gas cost and the execution time of running the \contract smart contract, for training a five-layer CNN for MNIST Dataset~\cite{mnist}. We split the CNN model such that the public model contains 5701 parameters, which are 15\% of the parameters in the entire model. Since an \eth block can accommodate a maximum of 250 parameters, we distribute the task of aggregating the public model into multiple contracts. The following measurement results are taken from one of the contracts, and we expect similar numbers in all other contracts. 

\subsubsection{Gas consumption}
\label{Gasmea}

\begin{figure}[htbp]
  \centering
  \includegraphics[width=\linewidth]{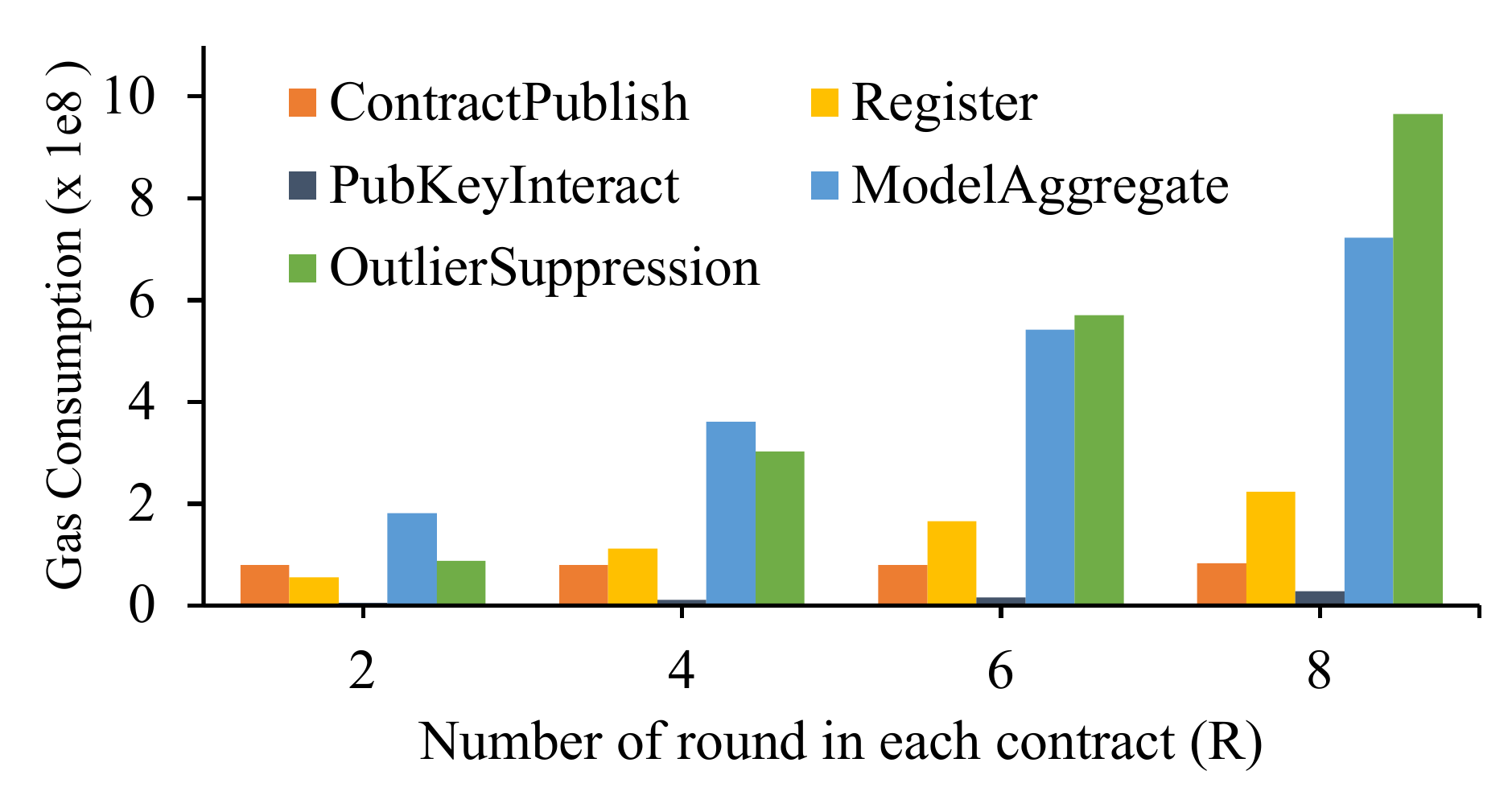}
  \vspace{-8mm}
  \caption{Gas consumption of \contract contract for $N=4$ data owners in each aggregation round and $\mu=25\%$ adversarial data owners, with different number of rounds.}
  \label{fig:impact_R_total}
\end{figure}

We fix the number of data owners in each round $N = 4$ and an estimated fraction of adversarial data owners $\mu = 25\%$, and evaluate the impact of number of groups $R$ on the gas consumption. As shown in~Figure~\ref{fig:impact_R_total}, for fixed $N$, the gas costs of \register, \PI, and \GA increase linearly with $R$. The computational complexity of multi-Krum scales quadratically with $R$, which leads to a faster increase in the gas cost of \OS.

\subsubsection{Execution time}

From the breakdown of the execution time in Tables~\ref{tbl:impact_R_time}, we observe that the overall contract execution time is dominated by the \GA and the \OS steps. As we increase the number of rounds $R$ for model aggregation, the execution time of the \OS step increases rapidly, leading to a significant increase in the overall execution time.

\begin{table}[htbp]
	\centering
	\small
	\caption{Breakdown of the \contract running time (ms) for different number of aggregation rounds.}
	\vspace{-2mm}
	\label{tbl:impact_R_time}
         \scalebox{0.9}{
	\begin{tabular}{|c|c|c|c|c|}  
        \hline
        & & & & \\[-6pt]
        $R$&2&4&6&8 \\
        \hline
        & & & & \\[-6pt]
        Register&1.75&1.85&1.77&1.3 \\
        \hline
        & & & & \\[-6pt]
        PubKeyInteract&0.53&0.58&0.68&0.71 \\
        \hline
        & & & & \\[-6pt]
        GradientAggregate&3.69&3.93&3.63&3.94 \\
        \hline
        & & & & \\[-6pt]
        OutlierSuppression&16.65&43.57&69.18&124.02 \\
        \hline
        & & & & \\[-6pt]
        Total&22.62&49.93&75.26&130.37 \\
        \hline
	\end{tabular}}
\vspace{-0.5cm}
\end{table}


\subsection{Model performance evaluations}
We now evaluate the model accuracy achieved by \scheme under various parameter choices. We assume that each $\DO$ has the same but a small amount of IID data. $\MO$ has an even smaller amount of IID data. 
We consider the task of training a five-layer CNN network for the MNIST dataset~\cite{mnist}. 
Parameters for the following experiments are show in ~Table~\ref{tab:para}.
\newcommand{\tabincell}[2]{
\begin{tabular}{@{}#1@{}}#2\end{tabular}
}

\begin{table}[htbp]
\footnotesize	
\centering
\caption{\label{tab:para}Design parameters for a image classification task on \scheme}

\begin{tabular}{llc}

\toprule
Parameter &Explanation & Default\\
\midrule
$\MO PreEp$ & Number of $\MO$ pre-train epochs & 20 \\
$\MO Ep$ & Number of $\MO$ local training epochs & 2 \\
$\DO Ep$& Number of $\DO$ local training epochs & 2 \\
$\DO Num$& Number of $\DO$s & 64 \\
 $N$&\tabincell{l}{Number of $\DO$s in each round of\\ \contract}  & 4 \\
 $R$& Number of rounds of \contract & 8 \\
$PL$& Number of layers in public model & 3 \\
 $Frag$& \tabincell{l}{Fragmentation of $\DO$ \\sparticipating in each iteration} & 0.5 \\
\bottomrule
\end{tabular}
\vspace{-0.5cm}
\end{table}

\begin{figure}[htbp]
  \centering
  \includegraphics[width=\linewidth]{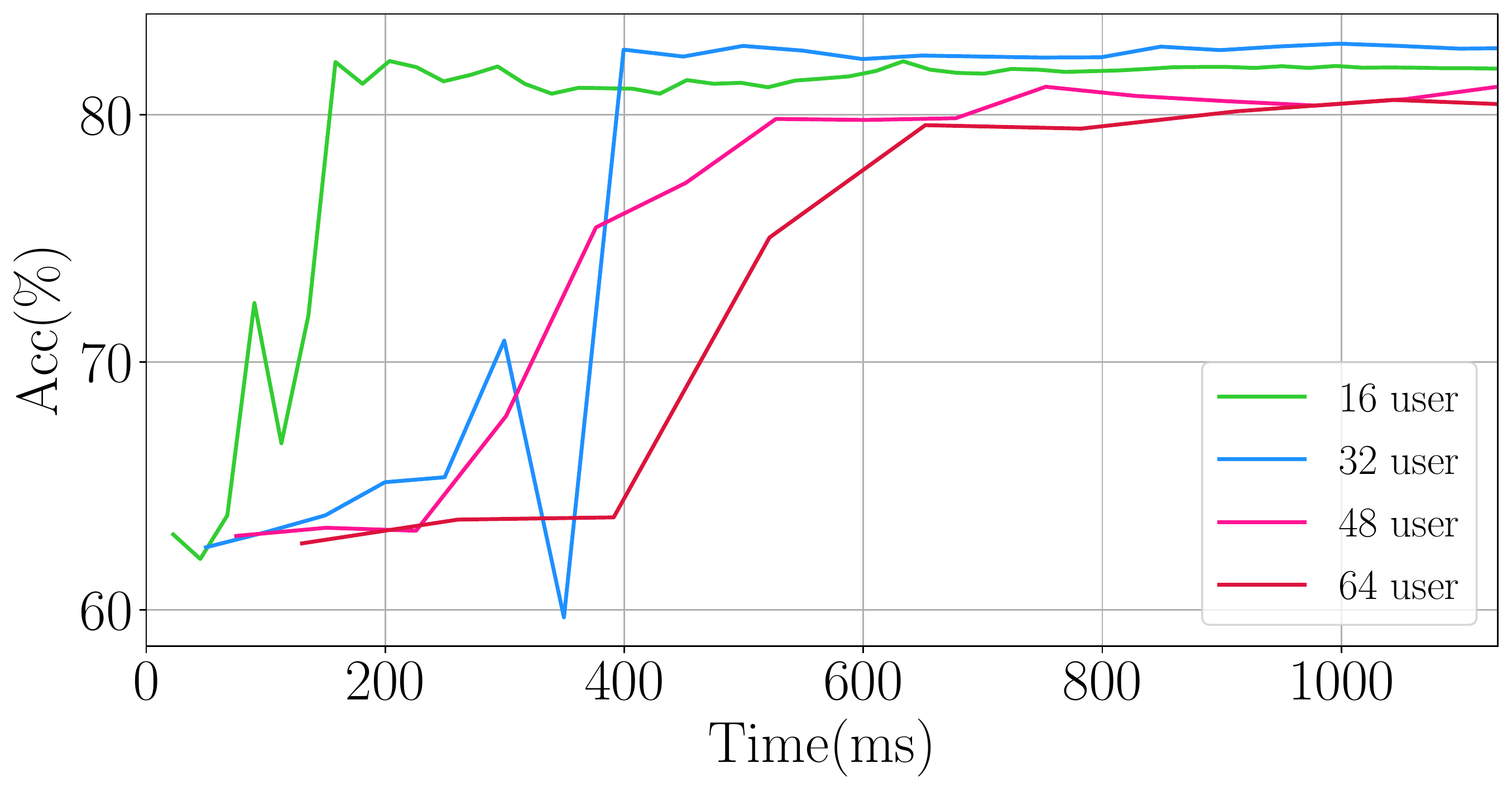} 
  \vspace{-8mm}
  
\caption{Accuracy for different number of data owners}
  \label{fig:impact_Users}
\end{figure}

To understand the impact of $\DO Num$ on the contract execution time and model accuracy, we fix $N=4$ and vary $\DO Num$. Consequently, the number rounds in \contract $R$ changes accordingly. When $\DO Num$ increases, the accuracy of the model can be improved faster, in terms of number of times the \contract contract is invoked. However, when the accuracy reaches a certain threshold, introducing more $\DO$s does not help any more.
We plot in Figure~\ref{fig:impact_Users} the model accuracy as a function of the total amount of blockchain processing time. While having more $\DO$s contributing to the model training allows the $\MO$ to invoke less number of \contract contract to reach certain model accuracy, the execution time of each contract is also longer. 

\begin{figure}[htbp]
  \centering
  \includegraphics[width=\linewidth]{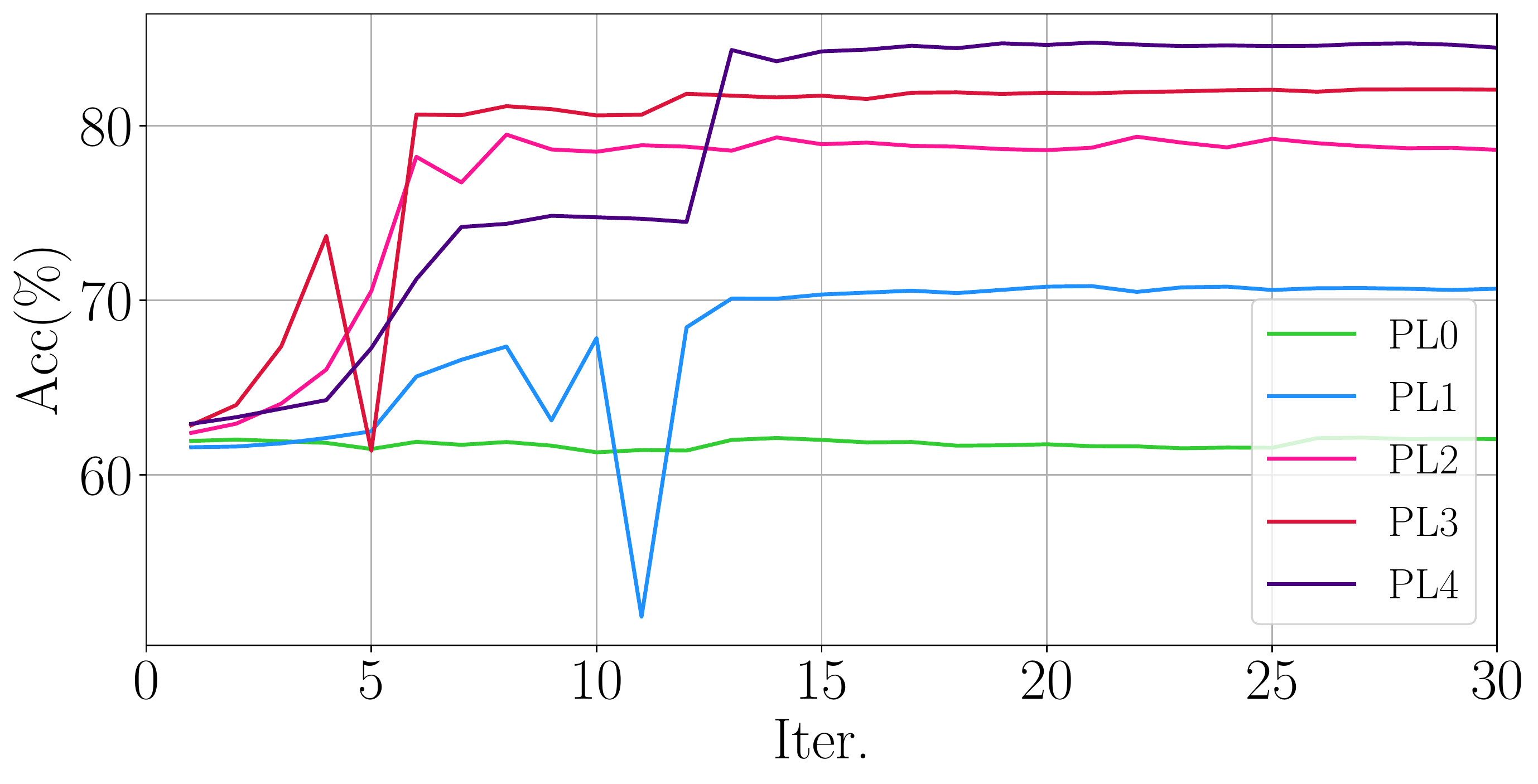} 
  \vspace{-8mm}
  \caption{Accuracy for different number of public layers}
  \label{fig:impact_PL}
\vspace{-0.4cm}
\end{figure}

Via changing $PL$, we examine the impact of the size and the structure of the public model on the training outcome. We consider different model splitting and public model configurations in Table~\ref{tbl:impact_Global_Layers}. When $PL = 0$, it means that $\MO$ only uses its own small dataset for local training. We take it as our baseline. After many iterations of training, it can only achieve 62\% accuracy. 
When $PL = 100\%$, it means that $\MO$ publishes all its model parameters on the blockchain.
We observe in Figure~\ref{fig:impact_PL} a sizable improvement for each increased layer in public model, demonstrating the effectiveness of \scheme on data crowdsourcing to help $\MO$'s learning task.


\renewcommand{\arraystretch}{1.2}
\begin{table}[htbp]
	\centering
	\caption{Considered public model configurations}
	\label{tbl:impact_Global_Layers}
         \scalebox{0.9}{
	\begin{tabular}{|c|c|c|}  
        \hline
        $PL$&Public layer&Public/Total \\
        \hline
        0&/&0\% \\
        \hline
        1&Linear 3&0.71\% \\
        \hline
        2&Linear 3+Linear 2&7.29\%  \\
        \hline
        3&Linear 3+Linear 2+ Conv 2&15.54\% \\
        \hline
        4&Linear 3+Linear 2+ Conv 2+ Conv 1&15.96\% \\
        \hline
	\end{tabular}}
\vspace{-0.5cm}
\end{table}

We also evaluate the influence of $\MO PreEp$, $\DO Ep$, and $\MO Ep$ on the model accuracy, and present the results in Table~\ref{tbl:impact_Epochs}. We observe that when $\MO PreEp$ is 80, the initial model is easy to be overfit, resulting in a lower accuracy of the final model. Choosing $40$ epochs of for pre-training gives the best model accuracy, over various combinations of local training epochs at the $\MO$ and $\DO$s.


\renewcommand{\arraystretch}{1.2}
\begin{table}[tp] 
\small
\centering
\caption{Impact of $Epochs$ (the 1st column is $\MO PreEp$).}  
\label{tbl:impact_Epochs}

\begin{tabular}{|c|c|c|c|c|c|c|}
\hline
\multirow{2}*{} & \multicolumn{3}{c|}{$\DO Ep = 1$} &  \multicolumn{3}{c|}{$\DO Ep = 2$} \\ 
\cline{2-7}
        & $\MO Ep = 1$ & $2$ & $4$ & $1$ & $2$ & $4$ \\
\hline 
20 &80.68 &81.37 & 82.46 &82.33& 70.74 &82.46\\
\cdashline{1-7}[0.8pt/2pt]
40  & 82.7  & 82.68  & 82.55 & 83.0& 82.83& 82.36\\
\cdashline{1-7}[0.8pt/2pt]
80  & 72.12  & 71.86  & 80.99& 82.19 & 73.28  & 73.1\\
\hline
\end{tabular}
\vspace{-0.5cm}
\end{table}

\subsection{Resilience on Byzantine data owners}

We simulate Byzantine data owners who instead of uploading local computing results to the contract, simply upload randomly generated data of the same dimension. We set the parameter $\mu$ in our outlier suppression mechanism to 0.25 and 0.5, then vary the actual number of Byzantine $\DO$s to observe the ability of \scheme
to defend Byzantine attacks. 
Through repeated experiments, we simulate different scenarios where different number of Byzantine $\DO$s join different rounds, and present the average accuracy of the final model in ~Table~\ref{tbl:impact_by}. We note that setting $\mu$ to be a higher value allows \scheme to be better resilient to attacks from Byzantine $\DO$s, but it will incur higher complexity in the contract execution, and increased running time and gas cost.

\begin{table}[htbp]
	\centering
	\caption{Average model accuracy achieved by \scheme under Byzantine data owners.}
	\label{tbl:impact_by}
         \scalebox{0.9}{
	\begin{tabular}{|c|c|c|}  
        \hline
        $Attack Rate$&$\mu=0.5$&$\mu=0.25$\\
        \hline
        2\%&83.12\%&76.62\% \\
        \hline
        4\%&82.89\%&79.72\%  \\
        \hline
        6\%&81.77\%&78.15\% \\
        \hline
        8\%&79.97\%&75.31\% \\
        \hline
        12\%&72.72\%&77.91\% \\
        \hline
        16\%&56.86\%&25.41\% \\
        \hline
      
	\end{tabular}}
\vspace{-0.5cm}
\end{table}
\section{Conclusion}
We develop \scheme, the first \eth smart contract implementation of a secure data market for decentralized machine learning. \scheme simultaneously achieves 1) model privacy against curious data owners; 2) data privacy against curious model and data owners; 3) resilience against Byzantine data owners who intend to corrupt model training; and 4) resilience to Byzantine model owner who tries to evade payment. We develop and deploy an \eth smart contract \contract, and measure its gas cost, execution time, and training performance over various system and design parameters. Through extensive experiments we observe high computation and cost efficiency of the contract, and high accuracy of the trained model, which demonstrate the applicability of \scheme as a practical secure data market.

\clearpage
\bibliography{reference}

\begin{thebibliography}{52}
\providecommand{\natexlab}[1]{#1}

\bibitem[{ETH(2021)}]{ETH}
 2021.
\newblock Ethereum smart contracts.
\newblock \url{https://ethereum.org/en/developers/docs/smart-contracts/}.
\newblock Accessed: 2021-06-05.

\bibitem[{get(2021)}]{geth}
 2021.
\newblock Official Go implementation of the Ethereum protocol.
\newblock \url{https://geth.ethereum.org/}.
\newblock Accessed: 2021-06-05.

\bibitem[{pyt(2021)}]{pytorch}
 2021.
\newblock Pytorch.
\newblock \url{https://pytorch.org/}.
\newblock Accessed: 2021-06-05.

\bibitem[{Rem(2021)}]{Remix}
 2021.
\newblock Remix - Ethereum IDE.
\newblock \url{https://remix.ethereum.org/}.
\newblock Accessed: 2021-06-05.

\bibitem[{Sol(2021)}]{Solidity}
 2021.
\newblock Solidity.
\newblock \url{https://docs.soliditylang.org/}.
\newblock Accessed: 2021-06-05.

\bibitem[{mni(2021)}]{mnist}
 2021.
\newblock THE MNIST DATABASE.
\newblock \url{http://yann.lecun.com/exdb/mnist/}.
\newblock Accessed: 2021-09-07.

\bibitem[{Web(2021)}]{Web3py}
 2021.
\newblock Web3py.
\newblock \url{https://web3py.readthedocs.io/en/stable/}.
\newblock Accessed: 2021-06-05.

\bibitem[{Al~Nuaimi et~al.(2015)Al~Nuaimi, Al~Neyadi, Mohamed, and
  Al-Jaroodi}]{al2015applications}
Al~Nuaimi, E.; Al~Neyadi, H.; Mohamed, N.; and Al-Jaroodi, J. 2015.
\newblock Applications of big data to smart cities.
\newblock \emph{Journal of Internet Services and Applications}, 6(1): 1--15.

\bibitem[{Bagdasaryan et~al.(2020)Bagdasaryan, Veit, Hua, Estrin, and
  Shmatikov}]{bagdasaryan2020backdoor}
Bagdasaryan, E.; Veit, A.; Hua, Y.; Estrin, D.; and Shmatikov, V. 2020.
\newblock How to backdoor federated learning.
\newblock In \emph{International Conference on Artificial Intelligence and
  Statistics}, 2938--2948. PMLR.

\bibitem[{Banerjee and Ruj(2018)}]{banerjee2018blockchain}
Banerjee, P.; and Ruj, S. 2018.
\newblock Blockchain Enabled Data Marketplace--Design and Challenges.
\newblock \emph{arXiv preprint arXiv:1811.11462}.

\bibitem[{Bell et~al.(2020)Bell, Bonawitz, Gasc{\'o}n, Lepoint, and
  Raykova}]{bell2020secure}
Bell, J.~H.; Bonawitz, K.~A.; Gasc{\'o}n, A.; Lepoint, T.; and Raykova, M.
  2020.
\newblock Secure single-server aggregation with (poly) logarithmic overhead.
\newblock In \emph{Proceedings of the 2020 ACM SIGSAC Conference on Computer
  and Communications Security}, 1253--1269.

\bibitem[{Bhagoji et~al.(2019)Bhagoji, Chakraborty, Mittal, and
  Calo}]{bhagoji2019analyzing}
Bhagoji, A.~N.; Chakraborty, S.; Mittal, P.; and Calo, S. 2019.
\newblock Analyzing federated learning through an adversarial lens.
\newblock In \emph{International Conference on Machine Learning}, 634--643.
  PMLR.

\bibitem[{Blanchard et~al.(2017{\natexlab{a}})Blanchard, El~Mhamdi, Guerraoui,
  and Stainer}]{blanchard2017machine}
Blanchard, P.; El~Mhamdi, E.~M.; Guerraoui, R.; and Stainer, J.
  2017{\natexlab{a}}.
\newblock Machine learning with adversaries: Byzantine tolerant gradient
  descent.
\newblock In \emph{Proceedings of the 31st International Conference on Neural
  Information Processing Systems}, 118--128.

\bibitem[{Blanchard et~al.(2017{\natexlab{b}})Blanchard, Mhamdi, Guerraoui, and
  Stainer}]{blanchard2017byzantine}
Blanchard, P.; Mhamdi, E. M.~E.; Guerraoui, R.; and Stainer, J.
  2017{\natexlab{b}}.
\newblock Byzantine-tolerant machine learning.
\newblock \emph{arXiv preprint arXiv:1703.02757}.

\bibitem[{Bommasani et~al.(2021)Bommasani, Hudson, Adeli, Altman, Arora, von
  Arx, Bernstein, Bohg, Bosselut, Brunskill
  et~al.}]{bommasani2021opportunities}
Bommasani, R.; Hudson, D.~A.; Adeli, E.; Altman, R.; Arora, S.; von Arx, S.;
  Bernstein, M.~S.; Bohg, J.; Bosselut, A.; Brunskill, E.; et~al. 2021.
\newblock On the Opportunities and Risks of Foundation Models.
\newblock \emph{arXiv preprint arXiv:2108.07258}.

\bibitem[{Bonawitz et~al.(2017)Bonawitz, Ivanov, Kreuter, Marcedone, McMahan,
  Patel, Ramage, Segal, and Seth}]{bonawitz2017practical}
Bonawitz, K.; Ivanov, V.; Kreuter, B.; Marcedone, A.; McMahan, H.~B.; Patel,
  S.; Ramage, D.; Segal, A.; and Seth, K. 2017.
\newblock Practical secure aggregation for privacy-preserving machine learning.
\newblock In \emph{proceedings of the 2017 ACM SIGSAC Conference on Computer
  and Communications Security}, 1175--1191.

\bibitem[{Chen, Su, and Xu(2017)}]{chen2017distributed}
Chen, Y.; Su, L.; and Xu, J. 2017.
\newblock Distributed statistical machine learning in adversarial settings:
  Byzantine gradient descent.
\newblock \emph{Proceedings of the ACM on Measurement and Analysis of Computing
  Systems}, 1(2): 1--25.

\bibitem[{Cornet and Holden(2018)}]{cornet2018systematic}
Cornet, V.~P.; and Holden, R.~J. 2018.
\newblock Systematic review of smartphone-based passive sensing for health and
  wellbeing.
\newblock \emph{Journal of biomedical informatics}, 77: 120--132.

\bibitem[{Data and Diggavi(2021)}]{data2021byzantine}
Data, D.; and Diggavi, S. 2021.
\newblock Byzantine-resilient high-dimensional SGD with local iterations on
  heterogeneous data.
\newblock In \emph{International Conference on Machine Learning}, 2478--2488.
  PMLR.

\bibitem[{Duan et~al.(2019)Duan, Zheng, Du, Zhou, Wang, and
  Au}]{duan2019aggregating}
Duan, H.; Zheng, Y.; Du, Y.; Zhou, A.; Wang, C.; and Au, M.~H. 2019.
\newblock Aggregating crowd wisdom via blockchain: A private, correct, and
  robust realization.
\newblock In \emph{2019 IEEE International Conference on Pervasive Computing
  and Communications (PerCom)}, 1--10. IEEE.

\bibitem[{Fang et~al.(2020)Fang, Cao, Jia, and Gong}]{fang2020local}
Fang, M.; Cao, X.; Jia, J.; and Gong, N. 2020.
\newblock Local model poisoning attacks to byzantine-robust federated learning.
\newblock In \emph{29th $\{$USENIX$\}$ Security Symposium ($\{$USENIX$\}$
  Security 20)}, 1605--1622.

\bibitem[{Geiping et~al.(2020)Geiping, Bauermeister, Dr{\"o}ge, and
  Moeller}]{geiping2020inverting}
Geiping, J.; Bauermeister, H.; Dr{\"o}ge, H.; and Moeller, M. 2020.
\newblock Inverting Gradients--How easy is it to break privacy in federated
  learning?
\newblock \emph{arXiv preprint arXiv:2003.14053}.

\bibitem[{Geyer, Klein, and Nabi(2017)}]{geyer2017differentially}
Geyer, R.~C.; Klein, T.; and Nabi, M. 2017.
\newblock Differentially private federated learning: A client level
  perspective.
\newblock \emph{arXiv preprint arXiv:1712.07557}.

\bibitem[{Ghosh et~al.(2020)Ghosh, Maity, Kadhe, Mazumdar, and
  Ramachandran}]{ghosh2020communication}
Ghosh, A.; Maity, R.~K.; Kadhe, S.; Mazumdar, A.; and Ramachandran, K. 2020.
\newblock Communication efficient and byzantine tolerant distributed learning.
\newblock In \emph{2020 IEEE International Symposium on Information Theory
  (ISIT)}, 2545--2550. IEEE.

\bibitem[{Hashem et~al.(2016)Hashem, Chang, Anuar, Adewole, Yaqoob, Gani,
  Ahmed, and Chiroma}]{hashem2016role}
Hashem, I. A.~T.; Chang, V.; Anuar, N.~B.; Adewole, K.; Yaqoob, I.; Gani, A.;
  Ahmed, E.; and Chiroma, H. 2016.
\newblock The role of big data in smart city.
\newblock \emph{International Journal of Information Management}, 36(5):
  748--758.

\bibitem[{Kim et~al.(2019)Kim, Park, Bennis, and Kim}]{kim2019blockchained}
Kim, H.; Park, J.; Bennis, M.; and Kim, S.-L. 2019.
\newblock Blockchained on-device federated learning.
\newblock \emph{IEEE Communications Letters}, 24(6): 1279--1283.

\bibitem[{Koutsos et~al.(2020)Koutsos, Papadopoulos, Chatzopoulos, Tarkoma, and
  Hui}]{koutsos2020agora}
Koutsos, V.; Papadopoulos, D.; Chatzopoulos, D.; Tarkoma, S.; and Hui, P. 2020.
\newblock Agora: A privacy-aware data marketplace.
\newblock In \emph{2020 IEEE 40th International Conference on Distributed
  Computing Systems (ICDCS)}, 1211--1212. IEEE.

\bibitem[{Krishnamachari et~al.(2018)Krishnamachari, Power, Kim, and
  Shahabi}]{krishnamachari2018i3}
Krishnamachari, B.; Power, J.; Kim, S.~H.; and Shahabi, C. 2018.
\newblock I3: An iot marketplace for smart communities.
\newblock In \emph{Proceedings of the 16th Annual International Conference on
  Mobile Systems, Applications, and Services}, 498--499.

\bibitem[{Li et~al.(2019)Li, Xu, Chen, Giannakis, and Ling}]{li2019rsa}
Li, L.; Xu, W.; Chen, T.; Giannakis, G.~B.; and Ling, Q. 2019.
\newblock RSA: Byzantine-robust stochastic aggregation methods for distributed
  learning from heterogeneous datasets.
\newblock In \emph{Proceedings of the AAAI Conference on Artificial
  Intelligence}, volume~33, 1544--1551.

\bibitem[{Liang et~al.(2020)Liang, Liu, Ziyin, Allen, Auerbach, Brent,
  Salakhutdinov, and Morency}]{liang2020think}
Liang, P.~P.; Liu, T.; Ziyin, L.; Allen, N.~B.; Auerbach, R.~P.; Brent, D.;
  Salakhutdinov, R.; and Morency, L.-P. 2020.
\newblock Think locally, act globally: Federated learning with local and global
  representations.
\newblock \emph{arXiv preprint arXiv:2001.01523}.

\bibitem[{Liu et~al.(2020)Liu, Ai, Sun, Zhang, Liu, and Yu}]{liu2020fedcoin}
Liu, Y.; Ai, Z.; Sun, S.; Zhang, S.; Liu, Z.; and Yu, H. 2020.
\newblock Fedcoin: A peer-to-peer payment system for federated learning.
\newblock In \emph{Federated Learning}, 125--138. Springer.

\bibitem[{Lyu et~al.(2020)Lyu, Yu, Nandakumar, Li, Ma, Jin, Yu, and
  Ng}]{lyu2020towards}
Lyu, L.; Yu, J.; Nandakumar, K.; Li, Y.; Ma, X.; Jin, J.; Yu, H.; and Ng, K.~S.
  2020.
\newblock Towards fair and privacy-preserving federated deep models.
\newblock \emph{IEEE Transactions on Parallel and Distributed Systems}, 31(11):
  2524--2541.

\bibitem[{Ma et~al.(2020)Ma, Li, Ding, Shi, Wang, Han, and
  Poor}]{ma2020federated}
Ma, C.; Li, J.; Ding, M.; Shi, L.; Wang, T.; Han, Z.; and Poor, H.~V. 2020.
\newblock When Federated Learning Meets Blockchain: A New Distributed Learning
  Paradigm.
\newblock \emph{arXiv preprint arXiv:2009.09338}.

\bibitem[{McMahan et~al.(2017)McMahan, Moore, Ramage, Hampson, and
  y~Arcas}]{mcmahan2017communication}
McMahan, B.; Moore, E.; Ramage, D.; Hampson, S.; and y~Arcas, B.~A. 2017.
\newblock Communication-efficient learning of deep networks from decentralized
  data.
\newblock In \emph{Artificial Intelligence and Statistics}, 1273--1282. PMLR.

\bibitem[{Mi{\v{s}}ura and {\v{Z}}agar(2016)}]{mivsura2016data}
Mi{\v{s}}ura, K.; and {\v{Z}}agar, M. 2016.
\newblock Data marketplace for Internet of Things.
\newblock In \emph{2016 International Conference on Smart Systems and
  Technologies (SST)}, 255--260. IEEE.

\bibitem[{Niu et~al.(2018{\natexlab{a}})Niu, Zheng, Wu, Gao, and
  Chen}]{niu2018achieving}
Niu, C.; Zheng, Z.; Wu, F.; Gao, X.; and Chen, G. 2018{\natexlab{a}}.
\newblock Achieving data truthfulness and privacy preservation in data markets.
\newblock \emph{IEEE Transactions on Knowledge and Data Engineering}, 31(1):
  105--119.

\bibitem[{Niu et~al.(2018{\natexlab{b}})Niu, Zheng, Wu, Tang, Gao, and
  Chen}]{niu2018unlocking}
Niu, C.; Zheng, Z.; Wu, F.; Tang, S.; Gao, X.; and Chen, G. 2018{\natexlab{b}}.
\newblock Unlocking the value of privacy: Trading aggregate statistics over
  private correlated data.
\newblock In \emph{Proceedings of the 24th ACM SIGKDD International Conference
  on Knowledge Discovery \& Data Mining}, 2031--2040.

\bibitem[{{\"O}zyilmaz, Do{\u{g}}an, and Yurdakul(2018)}]{ozyilmaz2018idmob}
{\"O}zyilmaz, K.~R.; Do{\u{g}}an, M.; and Yurdakul, A. 2018.
\newblock IDMoB: IoT data marketplace on blockchain.
\newblock In \emph{2018 crypto valley conference on blockchain technology
  (CVCBT)}, 11--19. IEEE.

\bibitem[{Raghupathi and Raghupathi(2014)}]{raghupathi2014big}
Raghupathi, W.; and Raghupathi, V. 2014.
\newblock Big data analytics in healthcare: promise and potential.
\newblock \emph{Health information science and systems}, 2(1): 1--10.

\bibitem[{Ramachandran, Radhakrishnan, and
  Krishnamachari(2018)}]{ramachandran2018towards}
Ramachandran, G.~S.; Radhakrishnan, R.; and Krishnamachari, B. 2018.
\newblock Towards a decentralized data marketplace for smart cities.
\newblock In \emph{2018 IEEE International Smart Cities Conference (ISC2)},
  1--8. IEEE.

\bibitem[{Shayan et~al.(2020)Shayan, Fung, Yoon, and
  Beschastnikh}]{shayan2020biscotti}
Shayan, M.; Fung, C.; Yoon, C.~J.; and Beschastnikh, I. 2020.
\newblock Biscotti: A Blockchain System for Private and Secure Federated
  Learning.
\newblock \emph{IEEE Transactions on Parallel and Distributed Systems}.

\bibitem[{Sheng et~al.(2013)Sheng, Tang, Xiao, and Xue}]{sheng2013sensing}
Sheng, X.; Tang, J.; Xiao, X.; and Xue, G. 2013.
\newblock Sensing as a service: Challenges, solutions and future directions.
\newblock \emph{IEEE Sensors journal}, 13(10): 3733--3741.

\bibitem[{So, G{\"u}ler, and Avestimehr(2021)}]{so2021turbo}
So, J.; G{\"u}ler, B.; and Avestimehr, A.~S. 2021.
\newblock Turbo-aggregate: Breaking the quadratic aggregation barrier in secure
  federated learning.
\newblock \emph{IEEE Journal on Selected Areas in Information Theory}, 2(1):
  479--489.

\bibitem[{Wang et~al.(2020)Wang, Sreenivasan, Rajput, Vishwakarma, Agarwal,
  Sohn, Lee, and Papailiopoulos}]{NEURIPS2020_backdoor}
Wang, H.; Sreenivasan, K.; Rajput, S.; Vishwakarma, H.; Agarwal, S.; Sohn,
  J.-y.; Lee, K.; and Papailiopoulos, D. 2020.
\newblock Attack of the Tails: Yes, You Really Can Backdoor Federated Learning.
\newblock In Larochelle, H.; Ranzato, M.; Hadsell, R.; Balcan, M.~F.; and Lin,
  H., eds., \emph{Advances in Neural Information Processing Systems},
  volume~33, 16070--16084.

\bibitem[{Wang, Kung, and Byrd(2018)}]{wang2018big}
Wang, Y.; Kung, L.; and Byrd, T.~A. 2018.
\newblock Big data analytics: Understanding its capabilities and potential
  benefits for healthcare organizations.
\newblock \emph{Technological Forecasting and Social Change}, 126: 3--13.

\bibitem[{Wang et~al.(2019)Wang, Song, Zhang, Song, Wang, and
  Qi}]{wang2019beyond}
Wang, Z.; Song, M.; Zhang, Z.; Song, Y.; Wang, Q.; and Qi, H. 2019.
\newblock Beyond inferring class representatives: User-level privacy leakage
  from federated learning.
\newblock In \emph{IEEE INFOCOM 2019-IEEE Conference on Computer
  Communications}, 2512--2520. IEEE.

\bibitem[{Wei et~al.(2020)Wei, Li, Ding, Ma, Yang, Farokhi, Jin, Quek, and
  Poor}]{wei2020federated}
Wei, K.; Li, J.; Ding, M.; Ma, C.; Yang, H.~H.; Farokhi, F.; Jin, S.; Quek,
  T.~Q.; and Poor, H.~V. 2020.
\newblock Federated learning with differential privacy: Algorithms and
  performance analysis.
\newblock \emph{IEEE Transactions on Information Forensics and Security}, 15:
  3454--3469.

\bibitem[{Yang et~al.(2019)Yang, Zhang, Fang, and Liu}]{yang2019byzantine}
Yang, H.; Zhang, X.; Fang, M.; and Liu, J. 2019.
\newblock Byzantine-resilient stochastic gradient descent for distributed
  learning: A lipschitz-inspired coordinate-wise median approach.
\newblock In \emph{2019 IEEE 58th Conference on Decision and Control (CDC)},
  5832--5837. IEEE.

\bibitem[{Yin et~al.(2018)Yin, Chen, Kannan, and Bartlett}]{yin2018byzantine}
Yin, D.; Chen, Y.; Kannan, R.; and Bartlett, P. 2018.
\newblock Byzantine-robust distributed learning: Towards optimal statistical
  rates.
\newblock In \emph{International Conference on Machine Learning}, 5650--5659.
  PMLR.

\bibitem[{Yin et~al.(2013)Yin, Sun, Cui, Hu, and Chen}]{yin2013lcars}
Yin, H.; Sun, Y.; Cui, B.; Hu, Z.; and Chen, L. 2013.
\newblock Lcars: a location-content-aware recommender system.
\newblock In \emph{Proceedings of the 19th ACM SIGKDD international conference
  on Knowledge discovery and data mining}, 221--229.

\bibitem[{Zhao et~al.(2020)Zhao, Zhao, Jiang, Tan, Niyato, Li, Lyu, and
  Liu}]{zhao2020privacy}
Zhao, Y.; Zhao, J.; Jiang, L.; Tan, R.; Niyato, D.; Li, Z.; Lyu, L.; and Liu,
  Y. 2020.
\newblock Privacy-preserving blockchain-based federated learning for IoT
  devices.
\newblock \emph{IEEE Internet of Things Journal}.

\bibitem[{Zhu and Han(2020)}]{zhu2020deep}
Zhu, L.; and Han, S. 2020.
\newblock Deep leakage from gradients.
\newblock In \emph{Federated learning}, 17--31. Springer.

\end{thebibliography}

\clearpage
\appendix
\section{Appendix A \\Pseudocode of $m$-Krum}
\begin{algorithm}[htbp]\label{alg:MK}
\caption{$m$-Krum}
\hspace*{0.02in} {\bf Input:}
$\{\vct{A}_r\}_{r \in \P}$: aggregation results of successfully executed rounds, $\mu$: fraction of rounds whose result are corrupted \\
\hspace*{0.02in} {\bf Output:}
{$\{\vct{A}_r\}_{r \in \P'}$ with $|\P'|=m$}
\begin{algorithmic}[1]\label{alg:MK}
\STATE ${\cal T} = \P$, $\P' = \emptyset$
\FOR{$i = 1,\ldots,m$}
\FOR{$r \in {\cal T}$}
\STATE neighbors $=|{\cal T}|-\mu|\P|-2$ closest ($\ell_2$ distance) vectors to $\vct{A}_r$
\STATE $S(r) = \sum_{\vct{A} \in \textup{neighbors}}||\vct{A}_r-\vct{A}||^2$\
\ENDFOR
\STATE $ r^*= \underset{r}{\argmin} S(r)$
\STATE $\cal T$.remove($r^*$)
\STATE $\P'$.add($r^*$)
\ENDFOR
\STATE return $\{\vct{A}_r\}_{r \in \P'}$
\end{algorithmic}
\end{algorithm}
\section{Appendix B \\ Additional Experiment Results}
\subsection{Smart contract measurements}

\subsubsection{Gas consumption}
\begin{figure}[htbp]
  \centering
  \includegraphics[width=\linewidth]{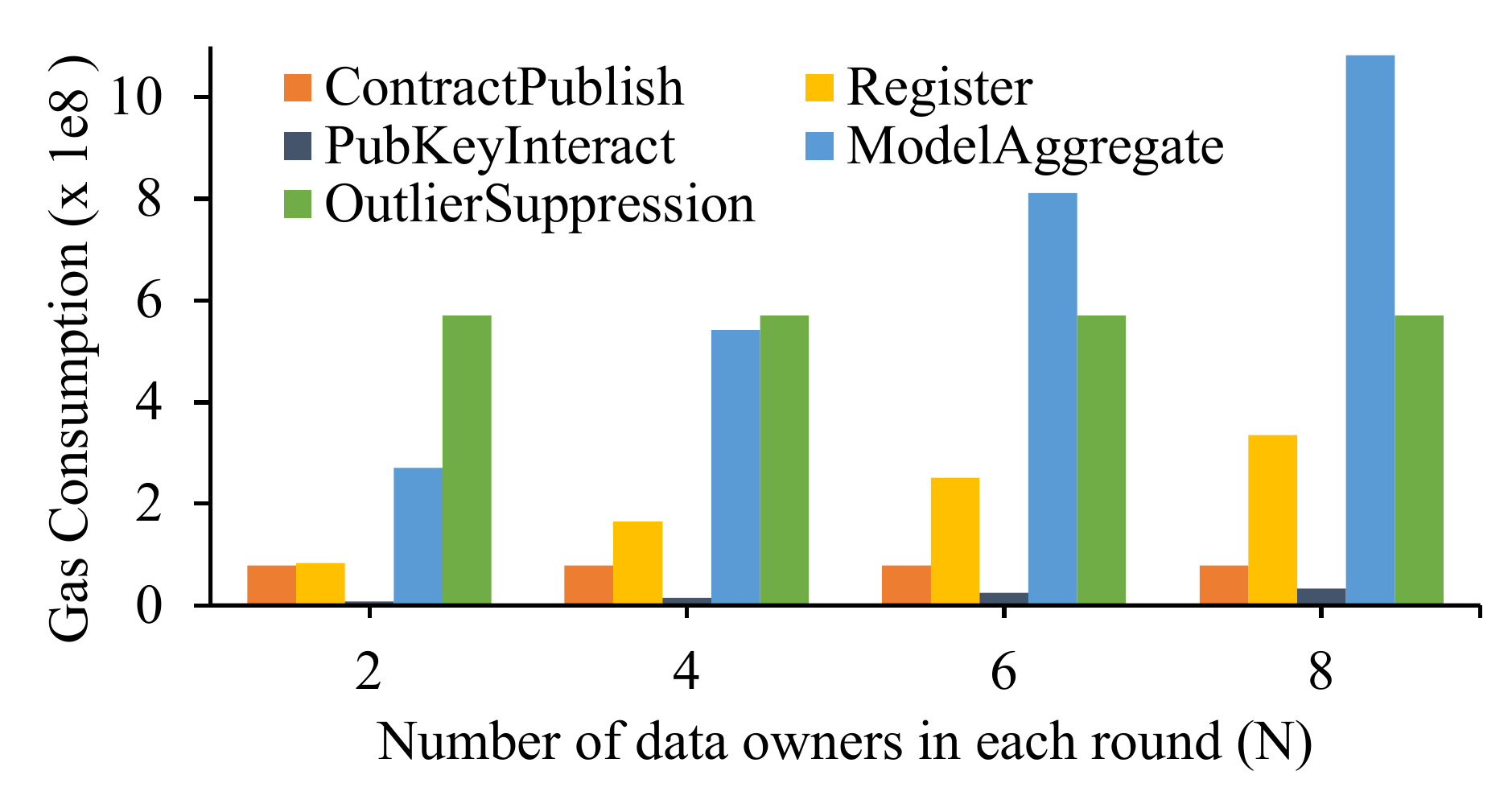} 
  \vspace{-8mm}
  \caption{Gas consumption of the \contract contract for $R=6$ aggregation rounds and $\mu=25\%$ adversarial data owners, for different number of data owners in each round.}
  \label{fig:impact_N_total}
 \end{figure}

We fix the number of \GA rounds $R=6$ and an estimated fraction of adversarial data owners $\mu = 25\%$, and measure the gas cost of the \contract contract for different number of $\DO$s in each round ($N$). We observe in Figure~\ref{fig:impact_N_total} that as $N$ varies from $2$ to $8$, the total gas consumption of \register, \PI, and \GA increases linearly $N$. In contrast, since the number of rounds $R$ has not changed, the computation complexity of the multi-Krum algorithm does not change, and the gas cost of \OS stays almost constant.

\vspace{-2mm}
 \vspace{-2mm}

\subsubsection{Execution time}
We present additional measurement results on the execution time of the \contract contract in Table~\ref{tbl:impact_N_time}. We note that as we increase the number of data owners in each round, the total execution time increases mildly as the execution time of the bottleneck operation \OS is not significantly affected by $N$. 

 \begin{table}[htbp]
            \centering
            \small
            	\vspace{-2mm}
            \caption{Breakdown of the \contract running time (ms) for different number of data owners in each round.}
            \label{tbl:impact_N_time}
         \scalebox{0.9}{
            
            \begin{tabular}{|c|c|c|c|c|}  
                \hline
                & & & & \\[-6pt]
                $N$&2&4&6&8 \\
                \hline
                & & & & \\[-6pt]
                Register&1.73&1.77&1.9&1.85 \\
                \hline
                & & & & \\[-6pt]
                PubKeyInteract&0.65&0.68&0.67&0.69 \\
                \hline
                & & & & \\[-6pt]
                GradientAggregate&3.43&3.63&3.65&3.85 \\
                \hline
                & & & & \\[-6pt]
                OutlierSuppression&67.78&69.18&72.53&73.43 \\
                \hline
                & & & & \\[-6pt]
                Total&73.59&75.26&78.76&79.81 \\
                \hline
            \end{tabular}}
            \vspace{-0.5cm}
        \end{table}

\end{document}